# A new physical simulation tool to predict the interface of dissimilar aluminum to steel welds performed by Friction Melt Bonding

T. Sapanathan[1,*], N. Jimenez-Mena[1], I. Sabirov[2], M.A. Monclús[2], J.M. Molina-Aldareguia[2], P. Xia[2,3], L. Zhao[1], A. Simar[1]

[1] Institute of Mechanics, Materials and Civil Engineering, UCLouvain, 1348 Louvain-la-Neuve, Belgium

[2] IMDEA Materials Institute, Calle Eric Kandel 2, 28906 Getafe, Madrid, Spain

[3] Universidad Politécnica de Madrid, E.T.S. de Ingenieros de Caminos, 28040 Madrid, Spain

* Corresponding author e-mail: thaneshan.sapanathan@uclouvain.be

**Abstract**

Optimization of the intermetallic layer thickness and the suppression of interfacial defects are key elements to improve the load bearing capacity of dissimilar joints. However, till date we do not have a systematic tool to investigate the dissimilar joints and the intermetallic properties produced by a welding condition. Friction Melt Bonding (FMB) is a recently developed technique for joining dissimilar metals that also does not exempt to these challenges. The FMB of DP980 and Al6061-T6 is investigated using a new physical simulation tool, based on Gleeble thermo-mechanical simulator, to understand the effect of individual parameter on the intermetallic formation. The proposed method demonstrates its capability in reproducing the intermetallic characteristics, including the thickness of intermetallic bonding layer, the morphology and texture of its constituents ($Fe_2Al_5$ and $Fe_4Al_{13}$), as well as their nanohardness and reduced modulus. The advantages of physical simulation tool can enable novel developing routes for the development of dissimilar metal joining processes and facilitate to reach the requiring load bearing capacity of the joints.

## 1. Introduction

Hybrid structures obtained via the interaction of chemically different constituents (i.e. dissimilar materials) have become attractive in modern engineering applications, since they combine the desirable properties of the constituents in one structure [1]. There are various methods for manufacturing hybrid structures [2, 3].

For processing metal/metal hybrid structures, welding offers strong financial and technical potential benefits including weight reduction, as it avoids intermediate components such as bolts, rivets or adhesives. Therefore, welding is highly preferred for joining dissimilar metallic materials into a hybrid structure. One of the most attractive hybrid structures for the automotive industry consists of aluminum and advanced high strength steel (AHSS), as their combination reduces the weight of the structure and maintains the high crashworthiness [4, 5]. However, due to the high reactivity between aluminum and steel in liquid state, there are limitations for the application of classical welding methods to join these materials. Early research studies showed that joining aluminum to steel using arc welding was not an appropriate approach due to the formation of thick and very brittle intermetallic compounds (IMC) in the molten pool of the bimetallic system [6]. This led to the further development of advanced techniques, which were adapted to keep one metal away from the other by adding a "bimetallic transition insert". These bimetallic inserts are comprised of one part of aluminum with equal part of steel already bonded to the aluminum by some other solid state welding, usually roll bonding, explosion welding or inertial friction welding [7-9]. More recently, researchers have also developed multi-stage procedures combined with arc welding (laser/arc welding) to join aluminum to steel, which are however time consuming and labor-intensive [9]. Moreover, laser welding-brazing [10] and laser key hole welding-brazing [11] enable to join aluminum to steel with the formation of various intermetallic compounds depending on the brazing material. Laser key-hole welding technique is also considered as a potential technique to join thick steel plates to aluminum alloys [12]. Although increase of laser beam penetration leads to defective joints, an optimum laser beam penetration can be used to potentially obtain strong welds [12].

Another approach was to retain the steel in solid state while aluminum was melted during the welding process. Recent advances in laser technologies enabled such new techniques for welding aluminum alloys to steels using an external magnetic field in laser welding or adding powders at the interface [13, 14]. During laser welding using the conductive mode, the steel plate is maintained in its solid state in order to minimize the reaction with the molten Al alloy. This method requires lower power compared to a conventional laser welding process [15, 16]. Thus, the diffusion of aluminum into steel at the interface is significantly reduced. This technology, therefore, allows reaching significantly higher welding speeds than in Friction Stir Welding (FSW) of aluminum to steel, up to several meters/min. The thickness of the Fe-Al intermetallic (IM) bonding layer may be controlled by the optimum combination of laser power and welding speed [16]. Nevertheless, some small manufacturing industries demand new welding techniques that are suitable to manufacture various hybrid structures, thus it is worth exploring other possibilities to fabricate hybrid structures by joining.

In this context, a novel welding process for joining dissimilar materials, Friction Melt Bonding (FMB), was recently developed and patented [17, 18]. Investigations have demonstrated that the weld quality dramatically depends on the FMB parameters, such as welding speed, rotational speed of the tool and exerted pressure [19]. Therefore, a significant number of experimental trials are required to determine the optimum FMB parameters resulting in high quality welds. There are two main factors controlling the quality of the weld: the IM bonding layer and the defects present in the weld [20, 21]. The thickness, morphology and, thus, properties of the IM bonding layer are determined by the thermal and pressure cycles seen by the interface during the FMB process. The critical defects from the viewpoint of the joint's strength are hot tears and shrinkage pores. As the present work focuses on the interface, the first factor (i.e. formation of IM layer during FMB process) will be studied in this manuscript, while the second factor (i.e. related to welding defects) is left for further investigations.

Morphology, thickness and properties of the IM layer play a key role in determining the bonding quality between steel and Al alloy [22-24]. There is a body of research focused on the growth kinetics of the IM layer at the steel/Al alloy interface. The outcomes of these activities may be summarized as follows [25-28]. Both static (i.e. steel plates are simply immersed inside the molten aluminum [29]) and dynamic (i.e. steel plates are placed and rotated inside the molten aluminum [30]) tests were conducted to investigate the kinetics of IM growth. The results from static and dynamic tests revealed that different IM phases were formed [25]. Formation of the IM layer during high power spot welding was studied in [31]. It was shown that the interdiffusion and, thus, IM formation was accelerated due to the high strain rate deformation in ultrasonic spot welding. The rate of interfacial reaction was over 6 times greater than the rate observed in diffusion couples using rate constants achieved from the static heat treatment condition. It was suggested that deformation-induced vacancies during the thermo-mechanical welding process accelerated the formation of the IM layer at the interface.

Analysis of experimental results have shown that the growth of the IM layers (FeAl, $Fe_3Al$, $Fe_2Al_5$) is controlled by diffusion in the Fe-Al bead weld following Eq. 1 [32-34].

$$l = Kt^{1/2} \qquad (1)$$

where, $l$ is the thickness, $K$ the temperature dependent constant and $t$ the exposure time. The growth of the $Fe_4Al_{13}$ IM layer is controlled by interface reaction and is linear with time [35]. The presence of alloying elements in steel can affect the growth of the IM bonding layer [35, 36]. For example, the temperature dependent constant $K$ decreased with increasing carbon content from 0.05 wt.% to 0.8 wt.% and saturated at high carbon content [36].

Very recently, a thermo-mechanical simulator was used to study the effect of temperature and time on the formation of $Fe_2Al_5$ ($\eta$ – phase) and $Fe_4Al_{13}$ ($\theta$ - phase)[1] and their growth kinetics at the interface between solid steel and molten Al [39]. However, the thermal cycles employed in [39] were not representative of the thermo-mechanical conditions in FMB. Moreover, Eq. 1 cannot be applied to predict the thickness of the IM bonding layer formed during the FMB process due to the non-linear temperature cycle seen by the Al/steel interface.

---

[1] It should be noted that the $\theta$ - phase was referred to as $FeAl_3$ in the earlier literature until the alternative form (i.e. $Fe_4Al_{13}$) was accepted by the research community after report by Grin *et al.* in 1994 [37]. (The alternative form ($Fe_4Al_{13}$) was originally found by Black in 1955 [38].

In this study, a new tool is designed for the physical simulation of the novel FMB process in order to reproduce the interface developed in a real dissimilar weld. The objective of the physical simulation tool is to predict the thickness, the microconstituents and the mechanical properties of the IM bonding layer in real welds produced by FMB as a function of the welding speed corresponding to the applied thermal cycle.

## 2. Materials and experimental procedures

### 2.1. FMB process and temperature measurements

FMB was performed in a lap welding configuration, with dual phase steel (DP980) plate placed on top of aluminum alloy (AA6061 –T6) plate, using a Hermle UWF 1001H universal milling machine. The chemical composition of both materials analyzed by Inductively Coupled Plasma (ICP) mass spectrometry are listed in Table 1. A schematic illustration of the FMB configuration is shown in Fig. 1. The rotating flat faced cylindrical tool made of cemented tungsten carbide (WC-Co) locally stirs the steel with a penetration depth of ~100 μm and generates sufficient heat to locally melt the aluminum underneath the tool. The tool has a backward tilt angle of 0.5° during the welding process. The FMB process was carried out with a rotational speed of 2000 RPM and two different welding speeds of 100 and 200 mm/min, so hereafter the studied samples will be referred to as "FMB100" and "FMB200", respectively.

Table 1: Chemical composition (wt. %) of Al alloy and steels used in both FMB and physical simulation.

| Alloying elements | Al | Fe | Mn | Si | Cu | Cr | Ni | C | Mg | Ti |
|---|---|---|---|---|---|---|---|---|---|---|
| AA6061 – T6 | 97.50 | 0.44 | 0.05 | 0.56 | 0.24 | 0.19 | | | 0.93 | |
| DP980 steel | 0.02 | 97.45 | 1.94 | 0.21 | 0.01 | 0.19 | | 0.16 | | 0.02 |
| S355J2G3C steel | 0.03 | 97.80 | 1.38 | 0.22 | 0.06 | 0.11 | 0.02 | 0.16 | | 0.01 |

The welded plates and the steel backing plate had dimensions of 250 mm x 80 mm (L × W) with a welded length of ~220 mm. The base materials of DP980 steel and AA6061 – T6 had thicknesses of 0.8 mm and 3.1 mm, respectively. A steel backing (support) plate with 5 mm thickness was also used in the experimental set up (Fig.1). A K-type thermocouple was inserted through the backing plate and placed at the interface between the backing plate and aluminum at the weld centerline (Fig. 1). The thermal cycle seen by the Al plate during FMB process was measured. The forces corresponding to the maximum pressure during FMB were measured using a rotating Kistler dynamometer, which was attached to the tool during FMB.

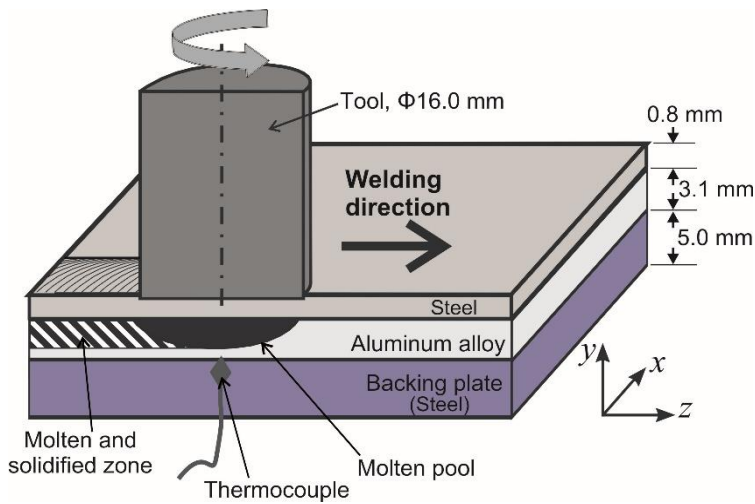

Fig. 1. A Schematic representation of the FMB process for the case of an aluminum/steel assembly with a backing (support) plate in lap welding configuration.

*2.2 Design of the physical simulation setup*

Physical simulation of the FMB process was carried out using a Gleeble 3800 (Dynamic Systems Inc, NY, USA) thermo-mechanical simulator. The system has a very high thermo-mechanical stability and has been widely used for physical simulation of various metallurgical processes [40]. A special experimental setup was designed for this physical simulation (Fig. 2). A cylindrical base (Fig. 2, left) and a cylindrical punch (Fig. 2, right) were machined from S355J2G3C steel. This steel was selected for the physical simulation since DP980 steel cannot be manufactured in the shape of large blocks. Nevertheless, both steels have a very similar chemical composition (Table 1). The cylindrical base contains a cavity with a diameter of 10 mm. An AA6061-T6 disk having a diameter of 10 mm and a thickness of 3 mm is placed inside the cavity, and the steel punch is inserted into the cavity until contact is made with the AA6061 disk. Thus, the obtained setup is symmetric, and its midsection (which is the symmetry plane) corresponds to the mid-section of the AA6061 disk. A K-type thermocouple was welded to the mid-section of the set-up. The heating/cooling experiments were carried out following the thermal plots measured during real FMB experiments with the weld speeds of 100 and 200 mm/min (Fig. 3a), so the thermal cycles seen by the interface were precisely applied in the Gleeble experiments. The temperature was controlled with an accuracy of ±1 °C/s.

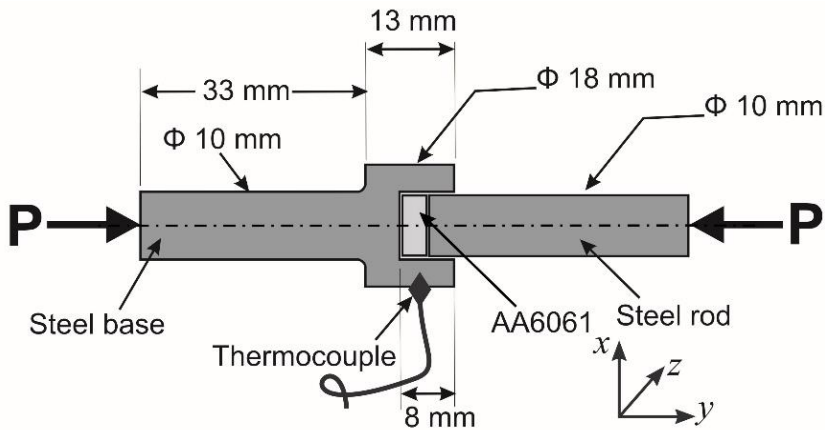

Fig. 2. Schematic drawing of the experimental setup developed for physical simulation of FMB process in the Gleeble 3800 thermo-mechanical simulator.

Pressure (P) was also varied with temperature during the physical simulation experiments. Initial holding pressure of 10 MPa was applied and then it was linearly increased to the set peak pressure and reached it simultaneously with the peak temperature. Then, the pressure was linearly reduced back to 10 MPa. For the samples corresponding to FMB100 and FMB200, maximum peak pressures of 30 and 35 MPa, respectively, were used during the Gleeble experiments (Fig. 3b). The pressure correspond to the vertical forces were measured using the dynamometer during the FMB process. After the melting of Al alloys, a pressure drop occurred during Gleeble experiment, which was compensated by the system. It should be noted that such an effect should also occur during real FMB process, when Al alloy is melted. After the Gleeble experiments, specimens were cross-sectioned along the longitudinal axis (y axis in Fig. 2) for a thorough microstructural analysis and mechanical characterization on the formed IM bonding layer. Hereafter, the specimens after physical simulation experiments will be referred to as "GB100" and "GB200".

*2.3 Microstructural and mechanical characterization of the interface*

Microstructural characterization was performed on the cross-section of the GB100 and GB200 samples and a transverse section of the FMB100 and FMB200 samples. Specimens for microstructural characterization were ground and polished to a mirror-like surface applying standard metallographic techniques with final polishing using OPS (colloidal silica). To study the morphology of the interface, a ZEISS FEGSEM Ultra 55 scanning electron microscope (SEM) was used, operating at an accelerating voltage of 15 kV in both secondary electron and backscatter electron modes. Electron backscatter diffraction (EBSD) analysis was conducted on the transverse section of the FMB100 sample and the corresponding GB100 sample. EBSD studies were performed using a FEI Quanta™ Helios NanoLab 600i, equipped with a NordlysNano detector controlled by the AZtec Oxford Instruments Nanoanalysis (version 2.4) software. The data were acquired at an accelerating voltage of 18 kV, a working distance of 8 mm, a tilt angle of 70° and a step size of 100 nm in a square scan grit. The orientation data were post-processed with HKL Post-processing Oxford Instruments Nanotechnology (version 5.1$^©$) software and TSL Data analysis version 7.3 software. Emphasis was laid on the analysis of the IM bonding layers formed between the steel and the Al alloy: identification of the microstructural constituents, their morphology, grain structure and crystallographic orientation.

Nanoindentation tests were performed with a Hysitron TI950 Triboindenter using a Berkovich diamond tip on the formed IM bonding layers. These layers were previously polished to achieve a smooth surface finish suitable for nanoindentation. In order to investigate possible mechanical property variations across the bond width, indentation rapid mapping and single quasistatic tests were performed on bond areas that were initially scanned using the inbuilt scanning probe microscopy (SPM) setup of the instrument. Single indentations were carried out in displacement control mode, at a constant strain rate ($\dot{\varepsilon} = \dot{h}/h$) of 0.1 s$^{-1}$, where $h$ is the penetration depth and $\dot{h}$ is the penetration rate of the indenter. At least 20 indents were performed on each bond, at an imposed maximum depth of 150 nm. Indentation maps were carried out in load control mode, using a maximum load of 5 mN and loading-hold-unloading times of 0.5 seconds. Maps of up to 484 (22x22) indents were performed with a separation between indents of 2.5 µm. The nanohardness and reduced elastic modulus were determined from the analysis of the load–displacement curves using the Oliver and Pharr method [41].

## 3. Results and Discussion

### 3.1 Thermal cycles measured during FMB process

Fig. 3 illustrates thermal cycles measured during FMB process for the two applied welding speeds. It is seen that an increase of welding speed from 100 mm/min to 200 mm/min shortens the thermal cycle time. The heating rate increases by nearly a factor of 2, while the 'soaking time' at the temperature above 650°C is reduced from 3.85 s to 0.94 s. Peak temperatures of 716 °C and 662 °C were produced for the case of welding with 100 and 200 mm/min speed, respectively (Fig. 3). These peak temperatures are above the melting temperature ($T_m$) of the used aluminum alloy, $T_m$ = 650 °C, which was measured by Differential Scanning Calorimetry (DSC) technique. Such significant difference in thermal cycles is expected to affect the morphology and properties of the bonding layer formed at the steel / Al alloy interface [21, 42].

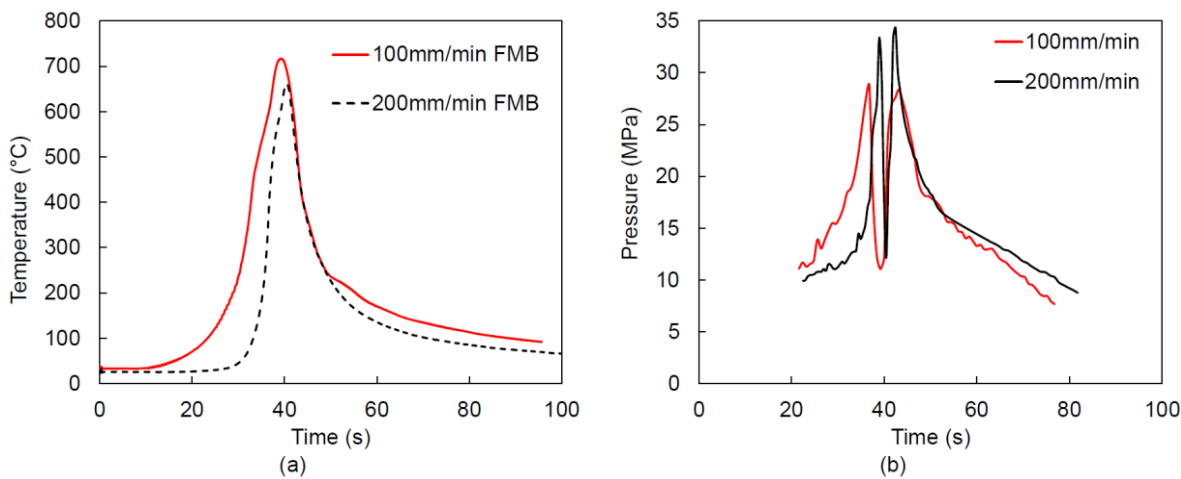

Fig. 3. (a) Thermal cycle recorded by thermocouple during FMB process with different welding speeds. These thermal cycles were exported into physical simulation of the FMB process using the Gleeble system. (b) The holding pressure curves obtained from Gleeble experiments showing the pressure simultaneously increases with the temperature and closely reaches the set values of 30 MPa and 35 MPa for GB100 and GB200, respectively.

### 3.2 Constituents and thickness of the IM layer

Fig. 4 shows SEM images of the IM bonding layers formed during both the FMB process with a welding speed of 100 mm/min (FMB100 sample) and the physical simulation using the same thermal cycle (GB100 sample). The top (in black color) part of images on Fig. 4a and b corresponds to the Al alloy, the bottom (in white color) to steel, and the area in-between (in gray color) is the IM bonding layer formed during the experiments. The IM layer exhibits a columnar microstructure in both samples (Fig. 4a and b), having similar average IM thicknesses of 38.4 µm in the FMB100 sample and 39.9 µm in the GB100 sample. It is also noticeable that the bonding layer has a sub-layer on top (in a slightly darker gray color indicated by ① in Fig. 4c and d) with an average thickness (based on surface average) of 3.0 µm in the FMB100 sample and 2.8 µm in the GB100 sample (Fig. 4c and d). In both samples, the top sub-layer has sharp edges. Thus, the thickness and morphology of these sub-layers are in good agreement between FMB and physical simulation. Microcracks are present in the interior of the IM bonding layer in both FMB100 and GB100 samples (indicated by black arrow in Fig. 4a and b). These microcracks result from the relaxation of diffusion induced stresses during cooling of the welded structure [43].

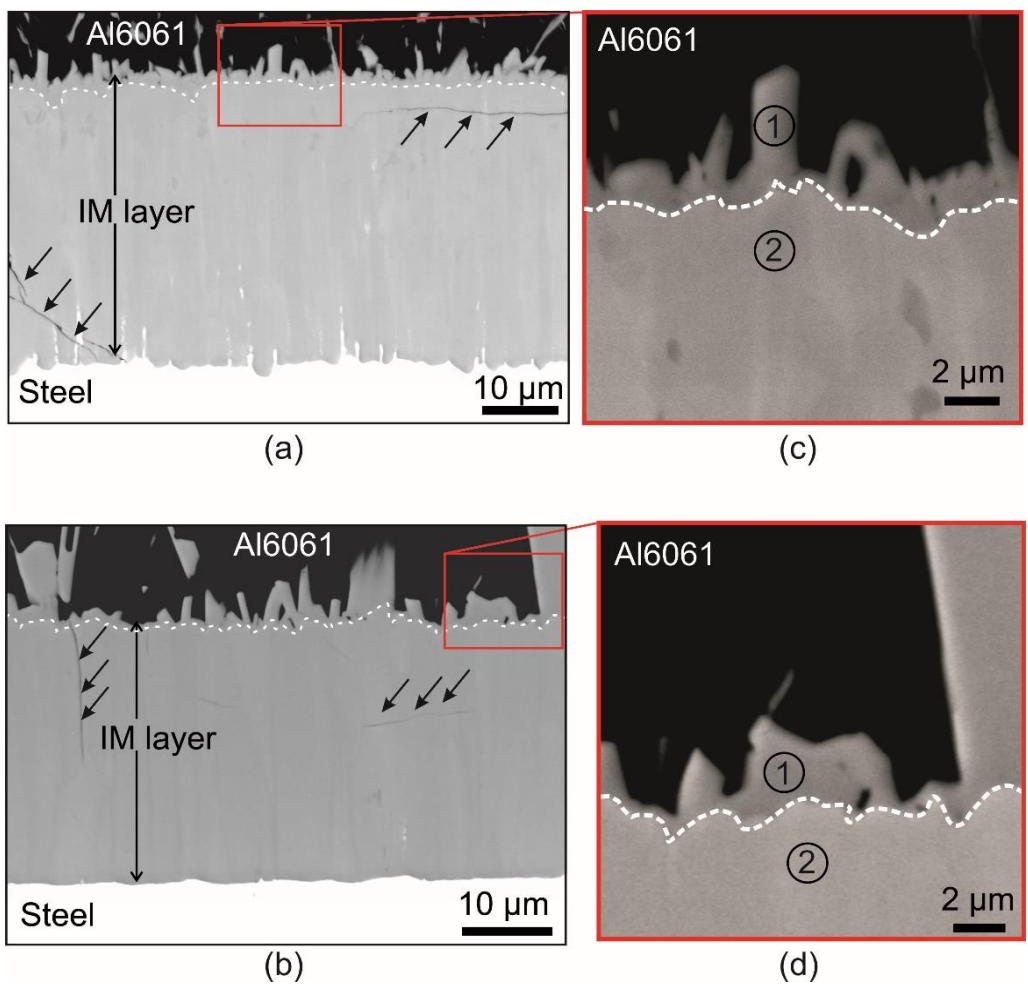

Fig. 4. Comparison of IM bonding layer thickness including the sub-layers of the IM revealing two distinct layers in dark and light grey in the transverse section of the (a) FMB100 sample and (b) physical simulation sample GB100. Morphology on the edge of the IM (aluminum side) showing sharp epitaxial growth for (c) FMB100 and (d) GB100. [① and ② are two different constituent layers of the IM bonding layer further investigated using EBSD phase mapping].

The IM bonding layers in FMB200 and corresponding GB200 samples are compared in Fig. 5. Again, a similar average thickness of the bonding layer was observed in both specimens. The FMB200 sample has a 7.4 µm thick bonding layer (Fig. 5a), while its thickness in the GB200 sample is 8.9 µm (Fig. 5b). These values are more than 4 times smaller than those measured for the bonding layers formed in the FMB100 and GB100 specimens. This observation can be rationalized based on the lower peak temperature and shorter thermal cycle measured for the welding with the speed of 200 mm/min (Fig. 3, Section 3.1). Indeed, both temperature and time are the key parameters determining interdiffusion of Fe and Al and the formation and growth of the bonding layer according to Fick's law [44] (see also Eq. 1). Both FMB200 and GB200 samples show a similar morphology (Fig. 5c and d) which is however different from that observed in the FMB100 and GB100 samples (Fig. 4c and d). The increase of welding speed (i.e. a change of thermal cycle) resulted in a blunt edge morphology of the IM bonding layer (Fig. 5c-d). Similar to the previous case, a sub-layer having a thickness of 2.5 µm and 2.3 µm can be identified on top of the bonding layer, indicated by ①, in the FMB200 (Fig. 5c) and GB200 (Fig. 5d) samples, respectively. Some microcracks are present in the interior of the bonding layer (Fig. 5), though their density is lower compared to that observed in the case of welding with 100 mm/min (Fig. 4). This can be related to the lower peak temperature and shorter thermal cycle (Fig. 3) which result in thinner IM bonding layer (Fig. 4 and Fig. 5) and lower level of diffusion induced residual stresses [43].

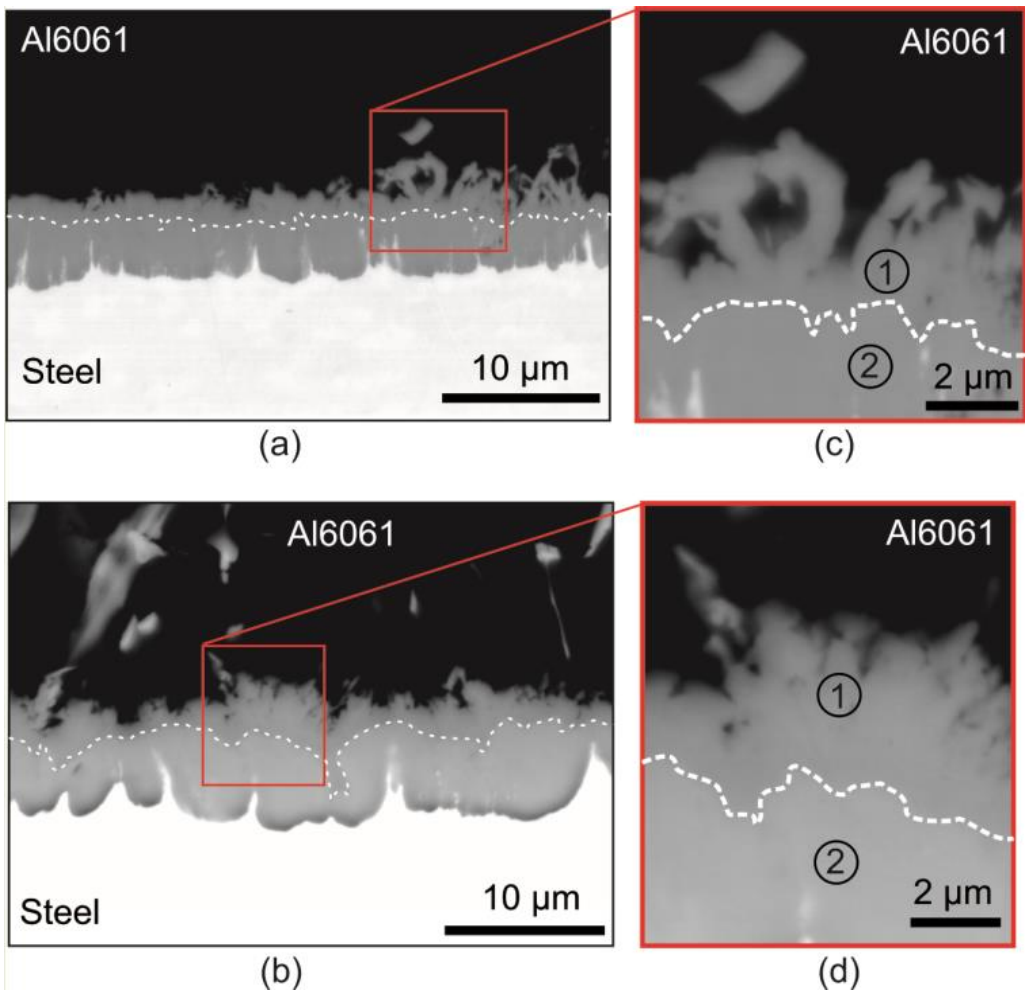

Fig. 5. Comparison of IM bonding layer for the transverse section of the (a) FMB200 sample and (b) physical simulation sample GB200. Morphology of IM constituent layers (dark and light grey) and on the edge of the intermetallic (aluminum side) revealing the epitaxial growth with blunt edges for (c) FMB200 and (d) GB200. [① and ② are two different constituent layers of the IM].

## 3.3 Phase composition and grain structure of the IM bonding layers

Our recent study on FMB welds identified that the main constituent is $Fe_2Al_5$ (η – phase, on steel side), whereas the top sub-layer presents $Fe_4Al_{13}$ (θ- phase, on aluminum side), using TEM-EDX (Transmission electron microscopy - Energy-dispersive X-ray spectroscopy) analysis combined with selected area electron diffraction (SAED) [24]. In this study, EBSD is used to analyze the crystal structure of the phase constituents of the IM bonding layer formed during real FMB process (FMB100) and physical simulation (GB100) are obtained as shown in Fig. 6. Two phases are identified in the bonding layer. Similar phases also observed for the FMB200 and GB200 samples (comparison provided in supplementary material Fig. S3). Thus, the microconstituents of the IM bonding layer can also be reproduced via physical simulation. However, a qualitative analysis of the grain structure formed in the bonding layer shows that the grains are much coarser in the GB100 and GB200 samples, though in both cases columnar grains are elongated in the direction of the IM layer growth. This effect can be related to the vibration of the steel/Al alloy structure during a real FMB process, which cannot be reproduced in the Gleeble thermo-mechanical simulator. Indeed vibration can influence the local stress field which is likely responsible for the development of sub structure during the growth of IM layer [45]. However, it is crucial to understand the formation mechanism of the IM layer to carefully deal with the welding and related issues. Based on the literature, one can suggest that the growth of the $Fe_4Al_{13}$ IM layer is controlled by interface reaction [35] while the growth of $Fe_2Al_5$ is governed by diffusion [32-34]. However, there are still open discussion on the subject of IM growth kinetics, and at present, we are working on this subject using in-situ experiments to reveal more insights on the growth kinetics.

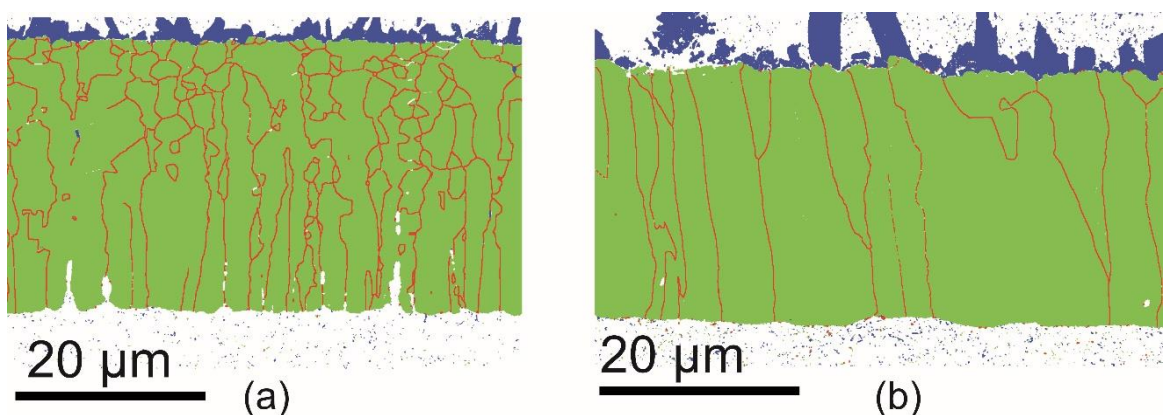

Fig. 6. Phase maps obtained from EBSD for samples: (a) FMB100 and (b) GB100. Blue and green represent $Fe_4Al_{13}$ and $Fe_2Al_5$, respectively. High angle grain boundaries (misorientation angle >15°) in $Fe_2Al_5$ phase are marked with red lines.

The inverse pole figure (IPF) maps of the corresponding areas of FMB100 and GB100 samples are presented in Fig. 7a-e. Both FMB and Gleeble samples show similar grain orientation distribution in all three orthogonal directions of $x, y, z$. Here, $x$ and $z$ directions are on the plane parallel to the interface while the direction $y$ is perpendicular to the interface. The IPF figures corresponding to the $x$ (Fig. 7a and b) and $z$ (Fig. 7e and f) directions show a slight texture in between <100> and <010> for the $Fe_2Al_5$ phase. Moreover, it can be noticed from Fig. 7c and Fig. 7d that the main microconstituent, i.e. $Fe_2Al_5$ phase, has a very strong <001> texture along the IM growth direction (i.e. $y$ direction perpendicular to the interface). Similar crystallographic textures were obtained for the FMB200 and GB200 samples (comparison provided in supplementary material Fig. S2). Based on the lattice parameter of Fe$_2$Al$_5$ (a=0.76559 nm along <100>, b=0.64154 nm along <010> and c=0.42184 nm along <001>) [46], the texture can be rationalized such that in the diffusion direction ($y$ direction perpendicular to the interface) the formation and growth of Fe$_2$Al$_5$ is much faster in the <001> direction than in other directions.

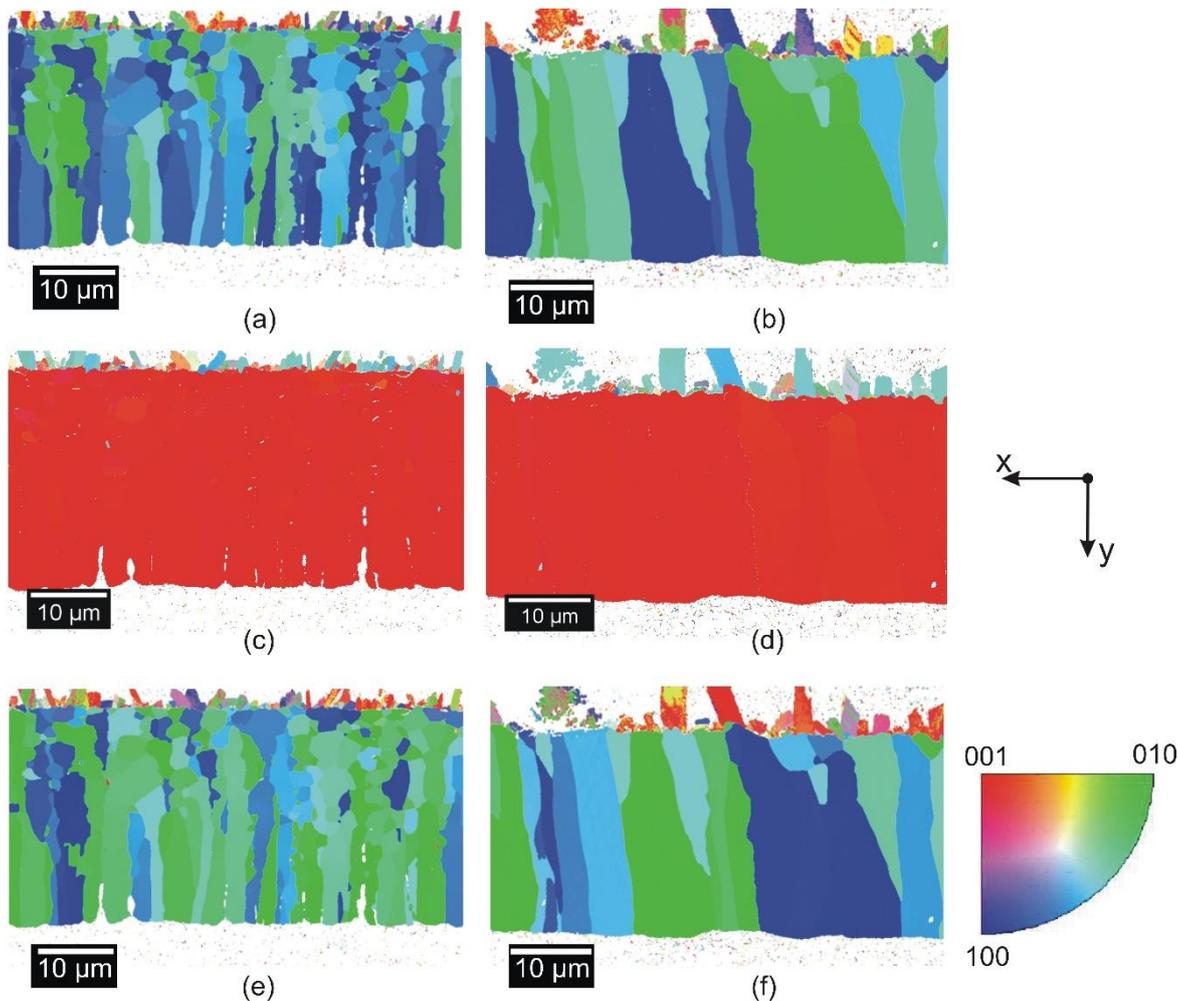

Fig. 7. IPF maps obtained along $x$ direction for (a) FMB100 and (b) GB100 samples. Strongly textured, <100>, IPF maps obtained along $y$ direction for (c) FMB100 and (d) GB100 samples. IPF maps obtained along $z$ direction for (e) FMB100 and (f) GB100 samples. The coordinate system used here is the same as the one used in Fig. 1 and 2.

*3.4 Hardness of the IM bonding layer*

A load of 5 mN was used to perform nanoindentation in an array fashion to obtain a map of hardness distribution across the interface. Representative load-displacement curves obtained for aluminum alloy, steel, IMC regions for both FMB and Gleeble samples are shown in Fig. 8. Good agreement can be noticed between the curves for FMB and Gleeble samples.

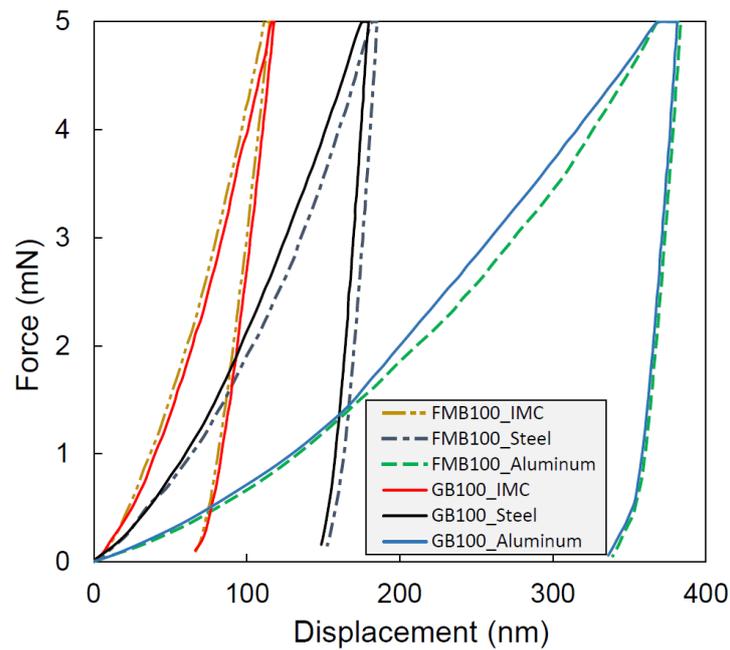

Fig. 8. Representative load-displacement curves obtained for aluminum alloy, steel, IMC from both FMB and Gleeble samples using a prescribed maximum load of 5 mN.

Fig. 9 shows optical microscopy images of the arrays of indents performed on the interface regions for both FMB100 and GB100 samples (Fig. 9a and b) and their corresponding nanohardness maps (Fig. 9c and d). The left side of the nanohardness maps indicated by light blue corresponds to steel, the right side (i.e. dark blue) corresponds to Al alloys and the central part is the IM bonding layer (Fig. 9c and d).

Overall, the nanohardness maps obtained for the interface region of both samples show a good agreement. Some 'soft spots' on the IM bonding layer and 'hard spots' on the areas corresponding to steel and Al alloys can be related to localized artifacts (i.e. roughness and/or contamination). Thin layers showing intermediate nanohardness between steel and IM bonding layer as well as between Al alloy and IM bonding layer appear due to local surface inclination causing uncertainty in the determination of the contact area of the indent. It should be noted that nanoindentation is generally sensitive to the surface preparation, and the studied "steel/IM layer/Al" alloy structure is complex for surface preparation due to significant differences in hardness of its constituents. A systematic nanoindentation was additionally performed along the lines parallel to the interface to study mechanical property variations in both samples (Fig. 10a and b). The quantitative results of the analysis are plotted in Fig. 10c.

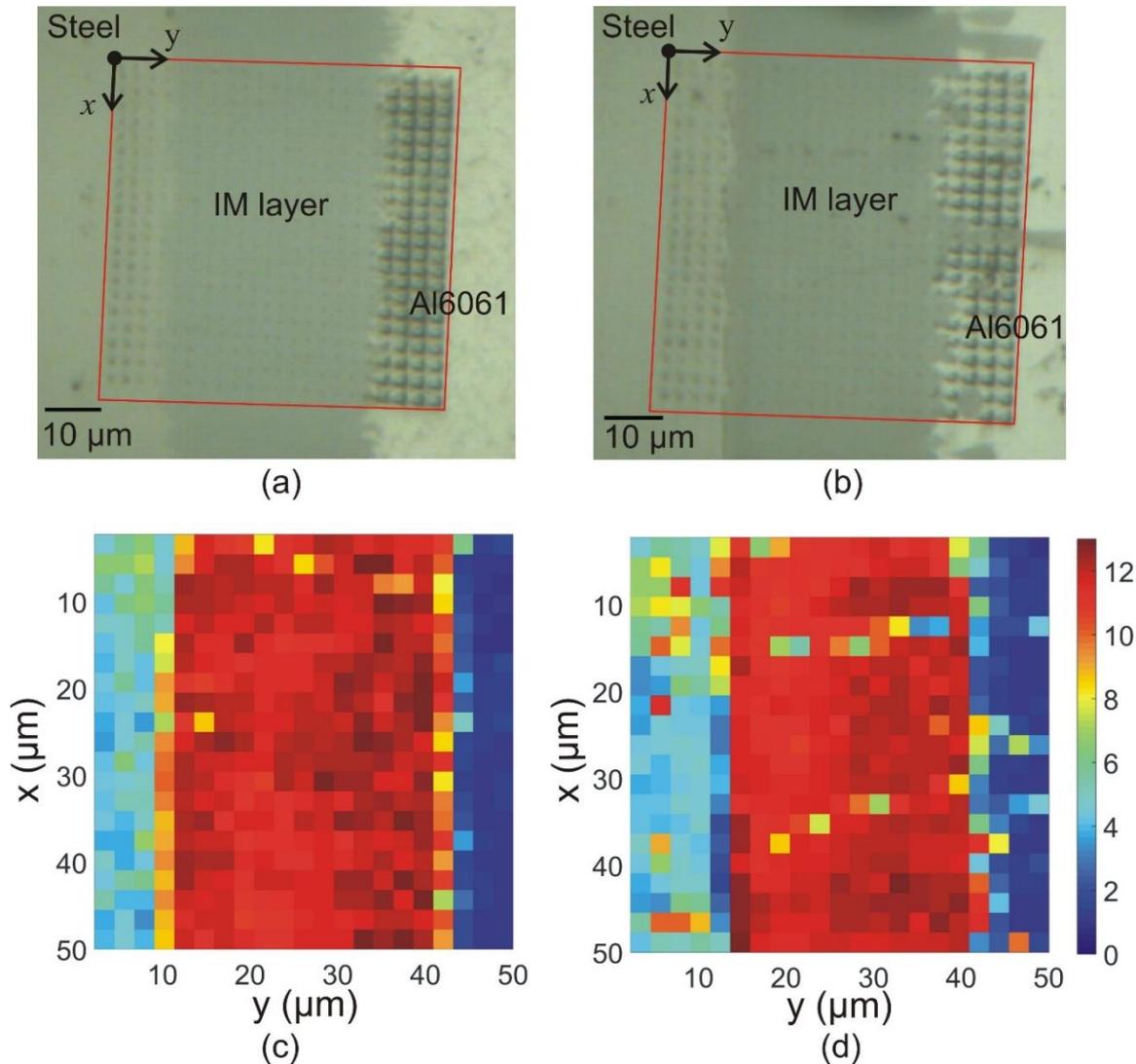

Fig. 9. Comparison of nanoindentation tests performed in arrays of 22 × 22 (484) points on the FMB100 and GB100 samples. Optical microcopy images showing the arrays of indents in the red box for (a) FMB100 and (b) GB100. Nanohardness distribution maps (in GPa) across the array for the interfaces of (c) FMB100 and (d) GB100 samples.

Those sampling points along lines parallel to the interface involve the η – phase of the IM bonding layer (Fig. 10a and b), which do not include the narrow region of the θ- phase. The measurements corresponding to lines 1-4 were obtained from the FMB100 while the lines 5-10 from GB100. The variations of nanohardness and reduced elastic modulus between the lines for a particular sample are negligible (Fig. 10). Moreover, the η – phase in the IM bonding layer of both FMB100 and GB100 samples presents nearly the same nanohardness and reduced elastic modulus. Nanohardness observed for η – phase of FMB200 and GB200 is also in good agreement (see supplementary material Fig. S4 for comparison). Thus, physical simulation is able to reproduce the IM bonding layer formed during the real FMB process extremely well, practically matching its mechanical properties. Based on the statistical analysis we can conclude that the η – phase of the IM produced by the thermo-mechanical conditions of FMB100 has a nanohardness of 12.4 ± 0.2 GPa and a reduced elastic modulus of 198.6 ± 3 GPa.

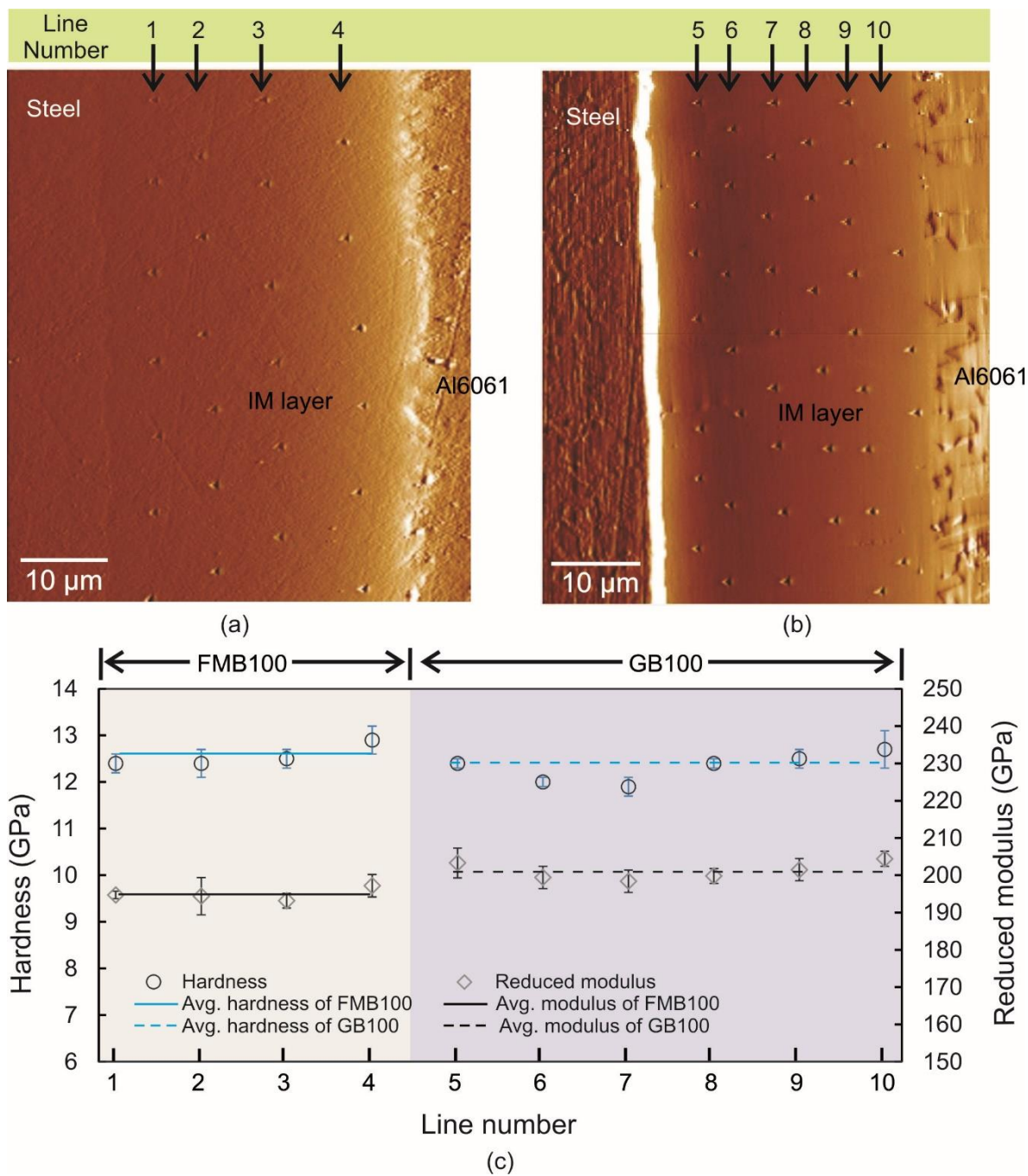

Fig. 10. Comparison of nanoindentation measurements performed along the parallel lines to the interface on IM bonding layers of the FMB100 and GB100 samples. Scanning probe microscopy (SPM) images of the indents showing (a) the indentation lines 1-4 corresponding to the FMB100 sample and (b) the indentation lines 5-10 corresponding to the GB100 sample. (c) Statistical average values of nanohardness (GPa) and reduced elastic modulus (GPa) along the lines (1-10) at the IM region come from both FMB100 and GB100 samples.

*3.5 Validity requirements on the outcome of the Gleeble experiments*

The results of this study clearly demonstrate that physical simulation can be successfully used to reproduce the thickness, phase composition (i.e. microconstituents), grain morphology, crystallographic texture and mechanical properties (nanohardness and reduced Young modulus) of the IM bonding layer formed between steel and Al alloy during experimental FMB process. Thickness, microstructure and properties of the IM bonding layer determine the bonding strength between two parts of the hybrid structure. It is clearly seen that these IM characteristics strongly depend on the FMB parameters. Therefore, selection of the optimum FMB parameters resulting in the optimum properties of the interface is of key importance for manufacturing high-quality hybrid structures. The trial-and-error approach based on the variation of processing parameters to determine the optimum ones requires significant amount of materials and large-scale experimental trials, which is costly and labor-intensive. Application of the developed physical simulation tool can greatly reduce the amount of materials and experimental trials required for determination of the optimum FMB parameters. The following procedure can be used to explore the optimum FMB parameters. First, theoretical models developed for the prediction of thermal cycles during FMB process [21] are applied to determine the potential ranges of the experimental cycles seen by the interface. In the next step, the calculated thermal cycles obtained from the theoretical models are used to establish matching parameters for the Gleeble experimental set-up and the Gleeble tests are performed. Finally, the analysis of the interfaces is carried out to determine the optimum thermal cycles. Once the optimum welding parameters are determined using a combination of the physical simulation and theoretical models, they can be readily transferred into the real FMB process.

4. **Conclusions**

A new physical simulation setup was proposed to reproduce a dissimilar Al/steel joint produced by Friction Melt Bonding (FMB). Thermal and loading conditions obtained from the FMB were applied as input to the physical simulation. It is demonstrated that the technique allows independently controlling the bonding parameters and reproducing the intermetallic layer thickness for the welding conditions of 100 and 200 mm/min. It also shows a good agreement on the IM growth morphology and the sub-layers of the intermetallic formed in the aluminum-steel system.

Physical simulation also reproduces the constituent layers of the IM. EBSD results confirm that the constituent layers mainly include η and θ phases (i.e. $Fe_2Al_5$ and $Fe_4Al_{13}$ respectively) of Al-Fe intermetallic compounds. Moreover, the IM grain growth and distribution are in good agreement for both FMB joints and Gleeble samples. Finally, nanoindentation results demonstrate that IM layers formed in FMB and Gleeble samples possess similar mechanical properties. These promising agreements in microstructural and mechanical behaviors foster future applications of the developed setup to investigate welding defects and to optimize the interface properties for various processing conditions.

Acknowledgements

NJM acknowledges the financial support of FRIA, Belgium. AS, LZ and TS acknowledge the financial support of the European Research Council for a starting grant under grant agreement 716678, ALUFIX project. PX acknowledges Chinese Scientific Council for financial support (No. 201606890031).

Supplementary material

Supplementary data associated with this article can be found, in the online version, at « <u>Please note, that the corresponding DOI link will be specified here later during the production</u> ».